\documentclass[aps,prb,twocolumn,groupedaddress,showpacs,floatfix,superscriptaddress,longbibliography]{revtex4-2}
 \usepackage[plainpages=false,pdfpagelabels,colorlinks=true,linkcolor=red,urlcolor=blue,citecolor=blue,pdftitle={Title},pdfauthor={},pdfdisplaydoctitle=true,pdfduplex=DuplexFlipLongEdge]{hyperref}
 \usepackage{graphicx}  
 \usepackage{dcolumn}   
 \usepackage[english]{babel}
 \usepackage{bm}        
 \usepackage{amsmath, amsthm, amssymb}
 \usepackage{bbold} 
 \usepackage{mathrsfs}
 \usepackage{array}
 \usepackage[usenames]{color}
 \usepackage{soul} 
 \usepackage{ulem}

\renewcommand\thesection{\arabic{section}}

\renewcommand\thesection{\arabic{section}}

\usepackage{changes}

\emergencystretch=\maxdimen
\begin{document}

\title{Order, Disorder and Monopole Confinement in the Spin-$1/2$ XXZ Model on a Pyrochlore Tube}
\author{Chunhan Feng}
\affiliation{Department of Physics, University of California, Davis, CA 95616, USA}
\author{Alexander Wietek}
\affiliation{Center for Computational Quantum Physics, Flatiron Institute, New York, New York 10010, USA}
\author{E. Miles Stoudenmire}
\affiliation{Center for Computational Quantum Physics, Flatiron Institute, New York, New York 10010, USA}
\author{Rajiv R. P. Singh}
\affiliation{Department of Physics, University of California, Davis, CA 95616, USA}
\date{\today}

\begin{abstract}
We study the ground state and thermodynamic properties of the spin-half XXZ model, with an Ising interaction $J_z$ and a transverse exchange interaction $J_{x}$, on a pyrochlore tube obtained by joining together elementary cubes in a one-dimensional array. Periodic boundary conditions in the transverse directions ensure that the bulk of the system consists of corner-sharing tetrahedra, with the same local geometry as the pyrochlore lattice. We use exact diagonalization, the density matrix renormalization group (DMRG), and minimally entangled typical thermal states (METTS) methods to study the system. When $J_z$ is antiferromagnetic ($J_{z}>0$) and $J_x$ is ferromagnetic ($J_{x}<0$), we find a transition from a spin liquid to an XY ferromagnet, which has power-law correlations at $T=0$. For $J_{z}<0$ and $J_{x}>0$, spin-two excitations are found to have lower energy than spin-one at the transition away from the fully polarized state, showing evidence for incipient spin-nematic order. When both interactions are antiferromagnetic, we find a non-degenerate ground state with no broken symmetries and a robust energy gap. 
The low energy spectra evolve smoothly from predominantly Ising to predominantly XY interactions.
In the spin-liquid regime of small $|J_{x}|$, we study the confinement of monopole-anti-monopole pairs and find that the confinement length scale is larger for $J_x<0$ than for $J_{x}>0$, although both length scales are very short. These results are consistent with a local spin-liquid phase for the Heisenberg antiferromagnet with no broken symmetries.

\end{abstract} 

\maketitle

\section{Introduction}

Recent years have seen significant advancements in computational techniques for quantum spin systems \cite{Schollwoeck2011}, allowing for a better understanding of their ground state phases including quantum spin liquids with long-range entanglement \cite{Savary16}. The extension of the density matrix renormalization group (DMRG) to cylinders of increasing width has allowed substantial progress to be made on two-dimensional quantum spin models such as the kagome lattice Heisenberg antiferromagnets\cite{Pollmann2017,White2011}. 

The study of three-dimensional quantum spin models are even more challenging as there are two transverse directions making the area-law entanglement grow very rapidly with increase in transverse dimensions. 

The quantum XXZ model on the pyrochlore lattice is a key model in the search for quantum spin liquid phases \cite{Canals00,Hermele04,Chern08,Benton18}.
The possibility that highly resonating quantum ground states can arise from the manifold of degenerate spin ice states, with exotic fractionalized quasiparticles and emergent gauge fields, has motivated many theoretical and experimental works \cite{Ramirez99,Gardner10,Gingras14,Rau19,Ross11,Bhardwaj21}. Yet, the ground state of perhaps the simplest such model, the Heisenberg antiferromagnet on the pyrochlore lattice, is not well established \cite{Shannon17,Benton18,Iqbal19,Schafer20,Yang21,Neupert21,Hagymasi21,Moessner98a,Okubo11,Benton15,Yuan16,Kazushi16,Kazuki19,Hering21,Berg03,Tsunetsugu01}, even as recent numerical studies observed a possible inversion symmetry broken ground state \cite{Hagymasi21,Neupert21}. 

In this paper we study the XXZ antiferromagnet on a pyrochlore tube, shown in Fig.~1. Periodic boundary conditions in the directions transverse to the tube imply that, in terms of local coordination, the system is the same as the pyrochlore lattice with corner sharing tetrahedron. We know that dimensionality plays a central role in the development of long-range order in the system. Here, our primary goal is to understand short-range behavior in the model, which is less sensitive to dimensionality and could be indicative of presence of short-range order in the full three-dimensional lattice as well. We also study ways in which the low dimensionality alters the long distance behavior.

\begin{figure}[hbpt]
\includegraphics[width=1\columnwidth]{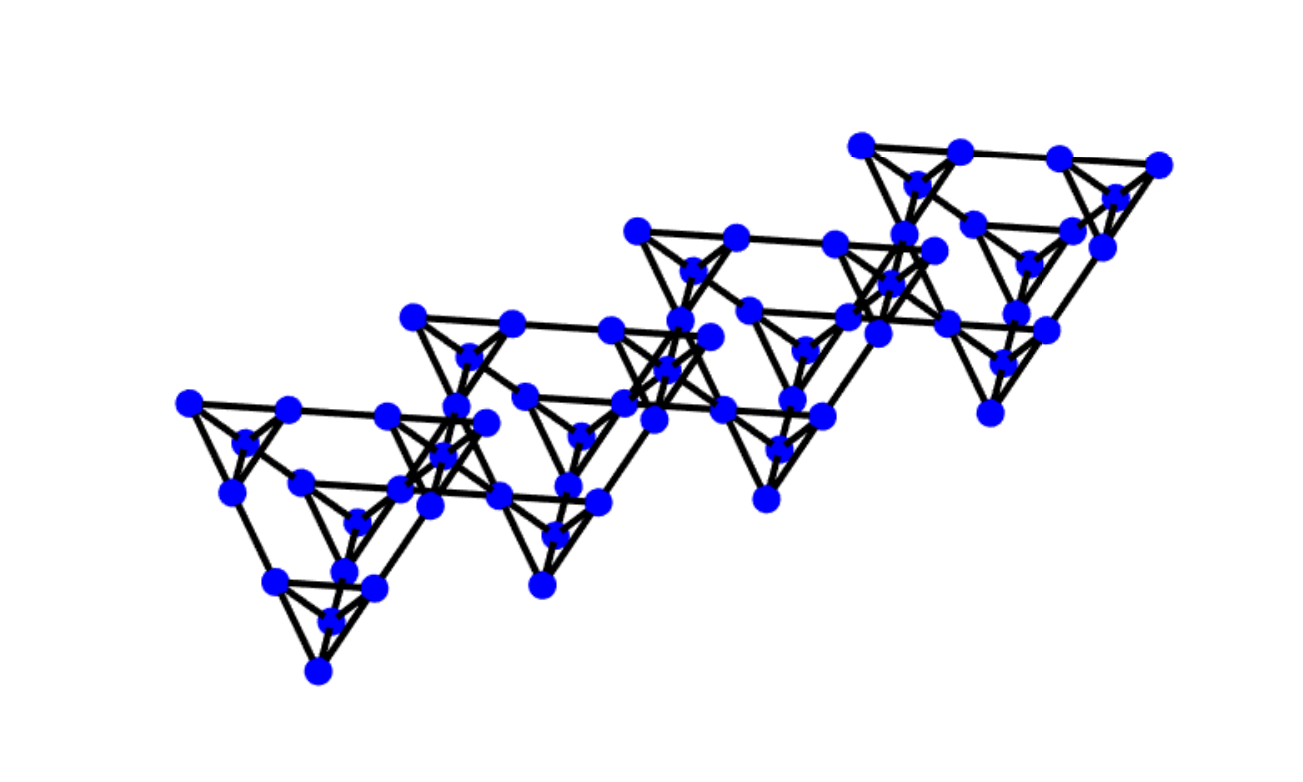}
\caption{A pyrochlore tube lattice with length $L=4$ (cubic unit cells) in the long direction. There are four tetrahedron in each cube and the total number of sites $N=4 \times 4 \times L =64$. }
\label{pyrochlore_lattice}
\end{figure}

\begin{figure}[hbpt]
\includegraphics[width=1\columnwidth]{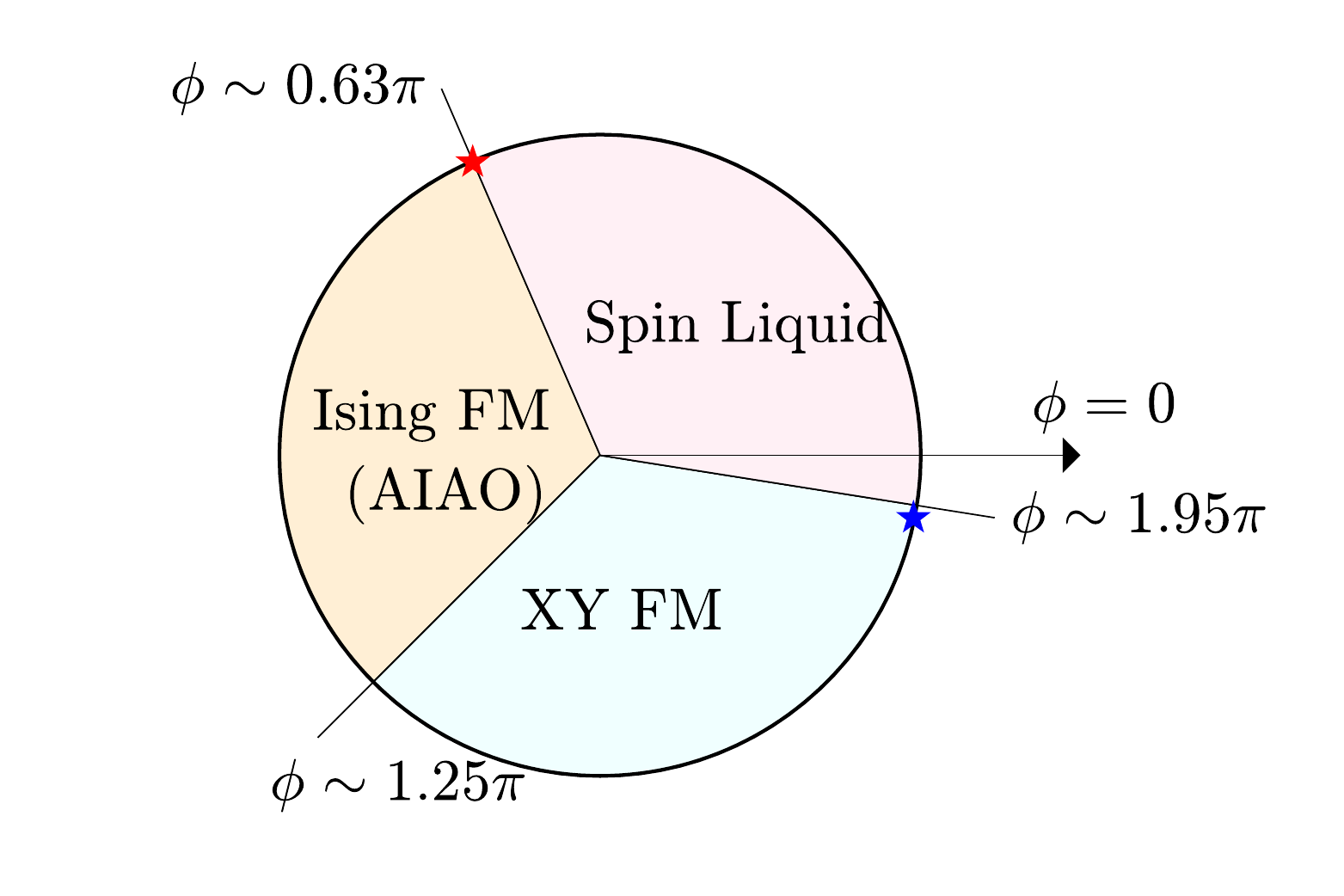}
\includegraphics[width=1\columnwidth]{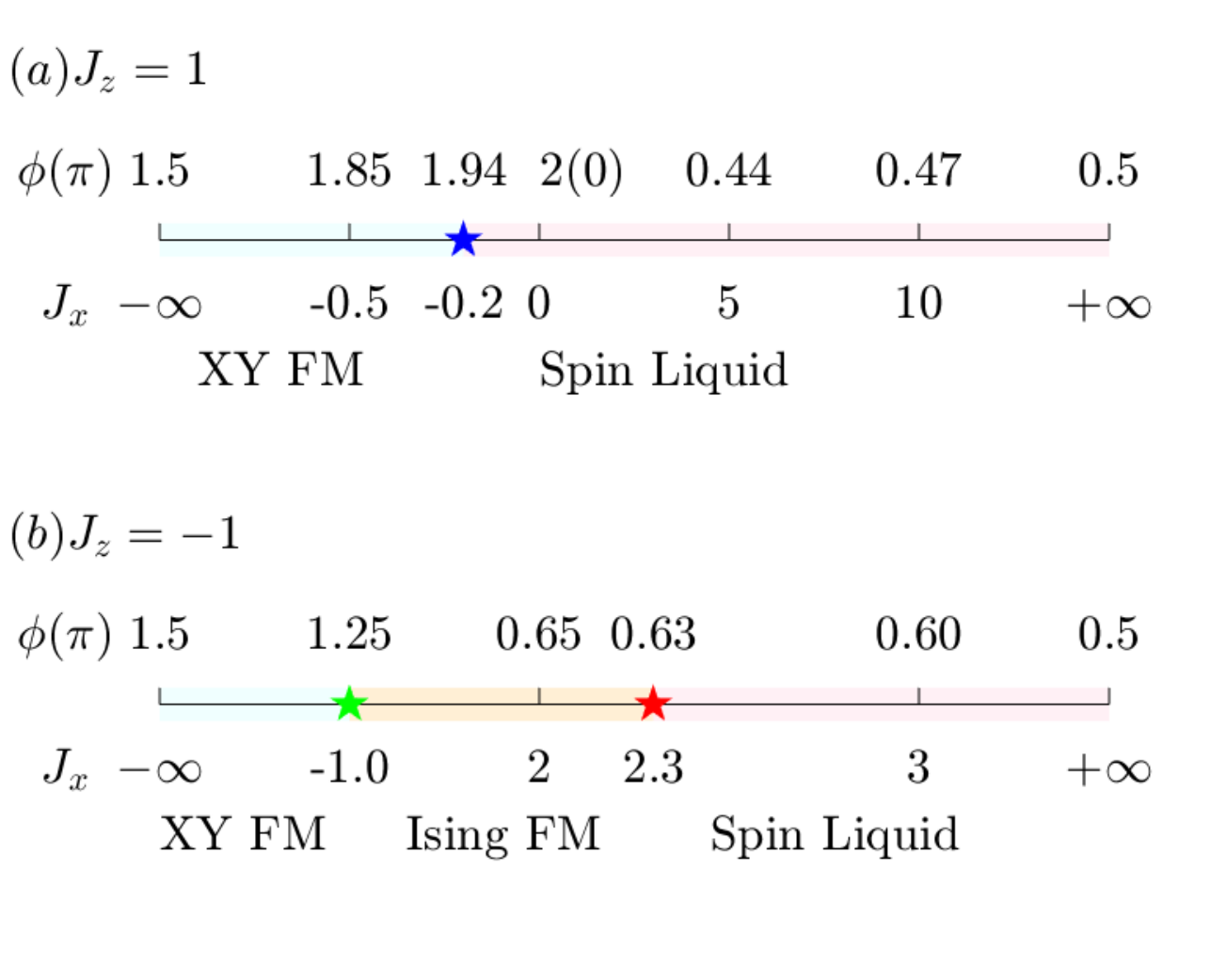}

\caption{
The upper panel shows the three primary ground state regimes of the model as a function of $\phi$ defined by the relations $J_z=J \cos{\phi}$, $J_x=J\sin{\phi}$: (i) Ising ferromagnet (FM) also called all-in-all-out (AIAO) phase (see text), (ii) XY ferromagnet and (iii) spin liquid. The nature of the spin liquid phase and whether there are multiple phases within the spin liquid region is a primary focus of our study. The lower panels show the phases with $|J_z| = 1$. }
\label{phasediagram}
\end{figure}

The three primary phases of the model are shown in Fig.~\ref{phasediagram}.
When the Ising coupling is antiferromagnetic and the XX coupling $J_x$ is ferromagnetic, our study finds a transition from the spin liquid phase to the XY ferromagnet as known for the pyrochlore lattice \cite{Kato2015}. There are quantitative differences in both the location of the transition point and the nature of the phases --- long-range order is replaced by power-law correlations due to one-dimensionality. Finite temperature properties at high and intermediate temperatures also agree between our system and the fully three-dimensional one \cite{Kato2015}. However, the low temperature thermodynamic properties are clearly different, as would be expected with changing dimensionality. 

Our main focus here is the case of antiferromagnetic transverse couplings. This case is particularly challenging because of the absence of quantum Monte Carlo algorithms without a sign problem for this system \cite{Kato2015, Motome06}. Previous studies have proposed a number of different types of order in the system including a variety of quantum spin liquids \cite{Benton15,Yang21, Shannon17} nematic quantum spin-liquid \cite{Benton18}, valence-bond order \cite{Hering21,Berg03}, broken symmetry between up and down pointing tetrahedron \cite{Hagymasi21,Neupert21} and broken time-reversal symmetry. Thus, even the question of short-range order in this system is far from settled. In our study, we find a non-degenerate ground state with no broken symmetries and a robust energy gap for the Heisenberg antiferromagnet. Our results support the development of short-range nematic correlations near the Heisenberg limit, which persists all the way to the XY limit ($J_{x}\to\infty$). In the latter case, such correlations can be seen by coming from the ferromagnetic Ising side \cite{Benton18}. As in the 3D case, we find that the lowest excitations in the ferromagnetic phase, on approach to the transition, carry spin-two and not spin-one, thus suggesting an incipient nematic order.

We also examine the confinement of monopole-antimonopole pairs due to the quasi-one dimensionality of the system in the spin ice phase at small transverse $J_x$. We look at the lowest energy state in the $S_z=1$ sector in a system with periodic boundary conditions. This forces at least two tetrahedra to not satisfy the ice rules. In other words, they contain the monopole excitations. For very small transverse couplings, there are only two such tetrahedra. We examine the distribution of distances between the monopole-antimonopole pairs. We find that for ferromagnetic transverse coupling the confinement length is larger than the confinement length for the antiferromagnetic transverse coupling although both length scales are quite short. 

Another issue of interest is the persistence of a finite temperature entropy plateaus in quantum
systems, where such plateaus must be rounded due to quantum fluctuations \cite{Applegate12,Kato2015}.
The minimally entangled typical thermal states (METTS) algorithm is an extension of DMRG that allows us to study the finite temperature properties of the system \cite{White09,Stoudenmire10}. We obtain heat capacity and entropy as a function of temperature. We find that the plateau in the entropy at the well-known Pauling value, a key signature of classical spin ice \cite{Ramirez99}, is lost as one moves away from the Ising limit. Ultimately, our quasi-one dimensional system has a robust gap and the temperature scale over which the entropy goes to zero from its spin-ice value is much larger than in the three-dimensional pyrochlore lattice.  This reflects the absence of a gapless photon mode in the quasi-one dimensional system. It is also consistent with the view that low temperature properties of the model are strongly modified due to the altered dimensionality.

\section{Model}


The XXZ Hamiltonian on a pyrochlore lattice is given by
\begin{align} 
 \mathcal{H} =  \sum_{\langle \mathbf{i}, \mathbf{j}\rangle} J_{z} S_{\mathbf{i}}^{z} S_{\mathbf{j}}^{z} + J_{x} (S_{\mathbf{i}}^{x} S_{\mathbf{j}}^{x}+S_{\mathbf{i}}^{y} S_{\mathbf{j}}^{y})
~~ ,
\label{eq:ham}
\end{align}
where $\langle \mathbf{i}, \mathbf{j}\rangle$ denotes the nearest neighbours on the pyrochlore lattice. $\Vec{S_{\mathbf{i}}}=\frac{\hbar}{2}\vec{\sigma_{\mathbf{i}}}$ (we set $\hbar=1$, and $\vec{\sigma_{\mathbf{i}}}$ are Pauli matrices) is the spin operator for site i. $J_x$ and $J_z$ describe the coupling strength in the transverse (X,Y) and Ising (Z) directions respectively. Both negative (ferromagnetic) and positive (antiferromagnetic) $J_z$ and $J_x$ will be considered in this work. To view the global phase diagram it is useful to define
\begin{equation}
    J_z= J\cos{\phi},\qquad J_x=J\sin{\phi},
\end{equation}
with $J>0$. For studying energy and temperature dependent properties, we set $|J_z|= 1$.

\begin{figure}[t]
\includegraphics[width=1\columnwidth]{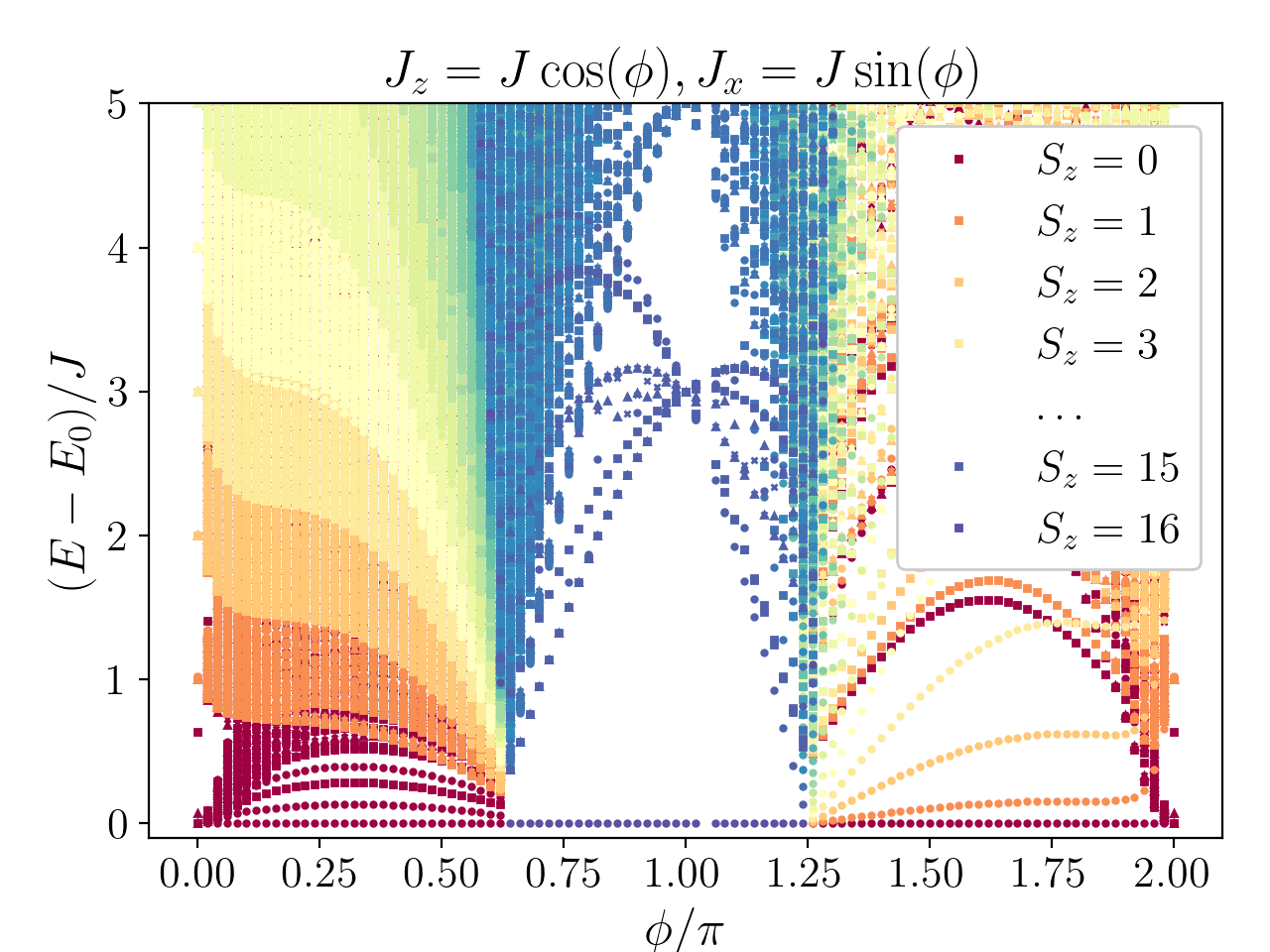}
\caption{
Low-lying spectrum of an $L=2$ periodic cluster with $32$ spins. The states are characterized by their $S_z$ quantum numbers. Three classes of ground-state regimes are evident from the figure. A high-$S_z$ ground-state phase in the middle, which corresponds to an Ising ferromagnet, and two $S_z=0$ ground-state regimes on the sides ending at the highly degenerate spin ice states at $\phi=0$ or $2\pi$. 
}
\label{spectrum}
\end{figure}

In physical realizations of the model in rare-earth pyrochlores, the spin-operators in Eq.~\ref{eq:ham} are defined in a local coordinate system, where the positive z-direction at a site points from the center of the neighboring down tetrahedron to the center of the neighboring up tetrahedron. Thus a ferromagnetic Ising order for our model in the $S_z$ direction corresponds to an All-In-All-Out (AIAO) state in the global coordinate system (Also shown in Fig.~\ref{phasediagram}), where all spins point in for each up tetrahedron and point out for each down tetrahedron or vice-versa. In this paper, we will mostly use the local coordinate system to describe the spin-state.

A pyrochlore tube lattice, consists of cubic unit cells of a face centered cubic (FCC) lattice with a four-point basis, joined along one direction. A tube of length $L=4$ (in terms of the cubic unit) ($N=64$ sites), is shown in Fig.~1. 
We use periodic boundary condition (PBC) in the transverse direction. In the DMRG and METTS studies, mostly open boundary condition (OBC) in the long direction is used to simplify the calculations.

A panoramic view of the phases of the model can be obtained from looking at the low-lying states of a $32$-site periodic cluster as shown in Fig~\ref{spectrum}. Each state is characterized by its conserved $S_z$ quantum number. The phase in the middle (near $\phi=\pi$) corresponds to the local-basis Ising ferromagnet or the  all-in-all-out (AIAO) phase. The phase to the right of this is a local XY ferromagnet, which ends in a quantum spin liquid as we approach the highly degenerate spin ice state at $\phi=2\pi$. The phase to the left of the AIAO phase is the primary focus of our investigation in this paper. It also ends in a quantum spin-liquid phase as we approach the highly degenerate spin ice state at $\phi= 0$.

\section{Methodology}

The density matrix renormalization group (DMRG) \cite{white92} has become the most powerful method for ground state studies of 1D systems. Here we apply it to a quasi-1D pyrochlore tube lattice. The important condition behind the success of the method is area-law entanglement and existence of a reduced state space which can capture all the interesting physics. 

DMRG in its modern form is based on a matrix product state (MPS) ansatz which is variationally optimized to converge to the desired ground state~\cite{Schollwoeck2011, Pollmann2017,White2011}. The ansatz is based on a Schmidt decomposition of wave functions. For any  bipartition of a system into subsystems $A$ and $B$, a wave function can be expressed as $|\psi \rangle=\sum_{i_A,j_B} M_{i_A,j_B} |i \rangle_A |j \rangle_B$ where $|i \rangle_A$ and $|j \rangle_B$ represent bases of states on subsystems $A$ and $B$. The matrix $M_{i_A,j_B}$ encodes the entanglement between the subsystems. Within a matrix product state approximation, one truncates the singular value decomposition $M =USV^\dagger$, by either choosing a maximal number of singular values (also referred to as the maximal bond dimension $D$) or by choosing $D$ such that the total sum of squares of truncated singular values, or truncated weight, is less than a cutoff $\varepsilon$. In our study, we typically use a truncated weight cutoff of $\varepsilon=10^{-6}$. We have checked the convergence of our results when decreasing the cutoff $\varepsilon$.

We use another MPS-based technique, the minimally entangled typical thermal states algorithm (METTS) \cite{White09,Stoudenmire10}, to compute finite temperature quantities. 
Instead of converting quantum problems into classical ones and sampling both quantum and thermal fluctuations as in a typical quantum Monte Carlo algorithm, METTS provides a way to directly sample quantum states. It begins with expressing the expectation value of an observable $\mathcal{O}$, at inverse temperature $\beta$ as 
\begin{align} 
\nonumber \mathcal{\langle O \rangle} 
&=  \frac{1}{\mathcal{Z}} \text{Tr} (e^{-\beta H}  \mathcal{O}) \\
&=  \frac{1}{\mathcal{Z}} \sum_{i} \langle i | e^{-\beta H/2} \mathcal{O} e^{-\beta H/2} | i \rangle \\
&= \frac{1}{\mathcal{Z}} \sum_{i} P(i) \langle \phi(i) \left|\mathcal{O}\right|\phi(i)\rangle
~~,
\label{eq:expectation}
\end{align}
where $\left|\phi(i)\rangle=P(i)^{-1/2}e^{-\beta H/2}\right|i\rangle$, and $P(i)=\langle i \left|e^{-\beta H}\right|i\rangle$. $\mathcal{Z}$ is the partition function and $|i\rangle$ is any orthonormal basis set. Notice that the probability of the quantum state $|\phi(i)\rangle$, $P(i)$, is real and non-negative. Thus, the METTS algorithm is free of any Monte Carlo sign problems as long as one can efficiently perform the imaginary time evolution of each state, which is the case for states of quasi-one-dimensional systems represented as MPS. Since the computational cost of using MPS increases rapidly as the entanglement entropy of a state grows, a natural choice of the orthonormal basis $| i \rangle$ is classical product states. The METTS algorithm allows the construction of a sequence of product states by collapsing the previous quantum state obtained after the operation of $e^{-\beta H/2}$ into a local basis. It guarantees that quantum states are sampled efficiently and with the desired distribution. 

The step which controls the efficiency and accuracy of the METTS algorithm is the operation of $e^{-\beta H/2}$ on each product state. This is first accomplished by the time-evolving block decimation (TEBD) \cite{Vidal2003,Vidal2004} algorithm for time-evolving MPS, which is controlled and accurate and allows one to adaptively grow the MPS bond dimension $D$. It is followed by the more efficient two-site time dependent variational principle (TDVP) \cite{Haegeman2011,Haegeman2016} algorithm (which scales as $ND^3d^2\beta$, where N is the number of sites and d is the degrees of freedom for the local site) to further increase the bond dimension $D$ and finally one-site TDVP for a fixed large bond dimension, which is approximately twice as fast as the two-site TDVP algorithm. The same approach was recently used to study finite temperature properties of the Hubbard model \cite{Wietek21,Wietek2021b}. More detailed discussions of these algorithms can be found in the review \cite{PAECKEL2019}. All simulations in this work use the ITensor library\cite{itensor}.

\section{Results for $T=0$ Properties}  
\label{sec:Results on Ground States}

In this section, we discuss DMRG results for ground state properties of the model. We compute the ground state energy, energy gap, bipartite entanglement entropy, and a variety of correlation functions of relevance to the different phases. For each set of Hamiltonian parameters, we typically perform thirty DMRG sweeps across the lattice. We set the maximum bond dimension to $D=2000$ and the truncation error cutoff to $\epsilon = 10^{-6}$, the latter setting the actual bond dimension used in the calculations. By varying $\epsilon$ we have checked that our results are well converged with respect to the truncation error.
We present results for different signs of $J_z$ and $J_{x}$ in the subsections below.

\subsection{Antiferromagnetic $J_z$ and Ferromagnetic $J_{x}$: Transition from a Spin Liquid to an XY Ferromagnet} 
\begin{figure}[hbpt]
\includegraphics[width=1\columnwidth]{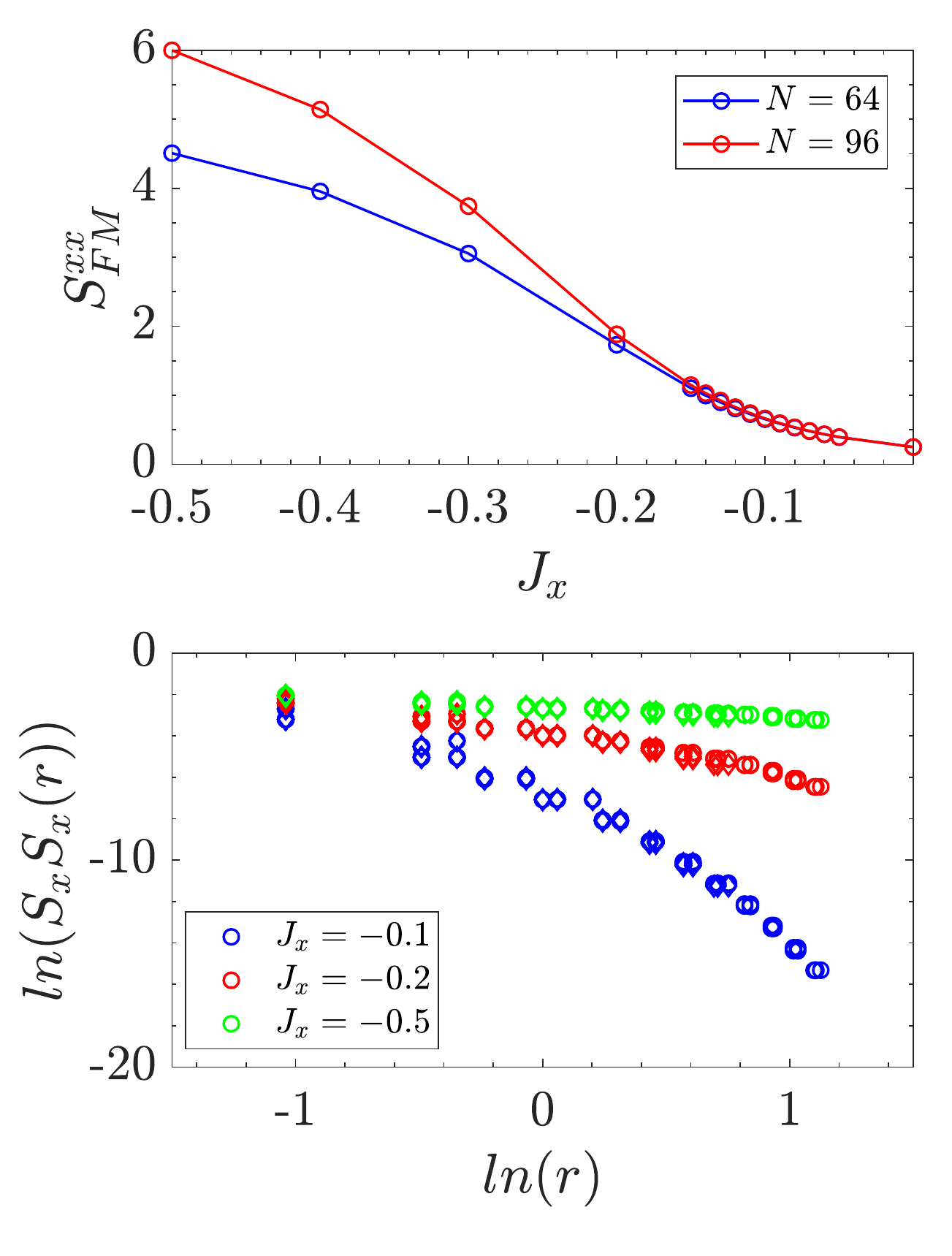}
\caption{Upper: Ferromagnetic structure factor in the $x$-component as a function of coupling strength $J_x$ for $N=64$ and $N=96$ systems for $J_z>0$ and $J_x < 0$. Lower: the logarithm of spin-spin correlations as a function of the logarithm of site distance. A linear relationship for $J_x=-0.2$ and $J_x=-0.5$ indicates a power lower decay of spin-spin correlations. $N=96$ and $N=64$ data are represented by circle and diamond symbols respectively. 
}
\label{fig:structure_factor_negativeJx}
\end{figure}
 
For ferromagnetic $J_x < 0$ and antiferromagnetic $J_z > 0$, the phase at small $|J_x|$ is a quantum spin-liquid which must be separated from an XY Ferromagnetic phase at larger $|J_x|$ by a phase transition. To study these phases and the transition, we calculate the transverse spin-spin correlation functions and structure factors
$S_{FM}^{xx}=S^{xx}(\mathbf{q}=0)$, where
\begin{equation}
    S^{xx}(\mathbf{q}) = \frac{1}{N} \sum_{\langle \mathbf{i}, \mathbf{j} \rangle } \langle S_{\mathbf{i}x} S_{\mathbf{j}x} \rangle e^{i\mathbf{q} \cdot (\mathbf{i}-\mathbf{j})} ,
\end{equation}
together with energy gap and entanglement entropy.
The ferromagnetic structure factor is plotted as a function of coupling strength $J_x$ in Fig.~\ref{fig:structure_factor_negativeJx} (a). Results for two lattice sizes $N=64$ and $N=96$ are shown in the plot. $S_{FM}^{xx}$ 
starts from a small value at $J_x=0$ and gradually increases as $J_x$ becomes more negative. When $\left| J_x\right|< \left| J_c \sim -0.2 \right|$, for different lattice sizes $S_{FM}^{xx}$ curves roughly overlap. In contrast, for $\left| J_x\right|> \left| J_c \right|$, there is a significant increase in $S_{FM}^{xx}$ as lattice size grows, indicating development of longer range ferromagnetic (FM) correlations in the transverse direction.

\begin{figure}[hbpt]
\includegraphics[width=1\columnwidth]{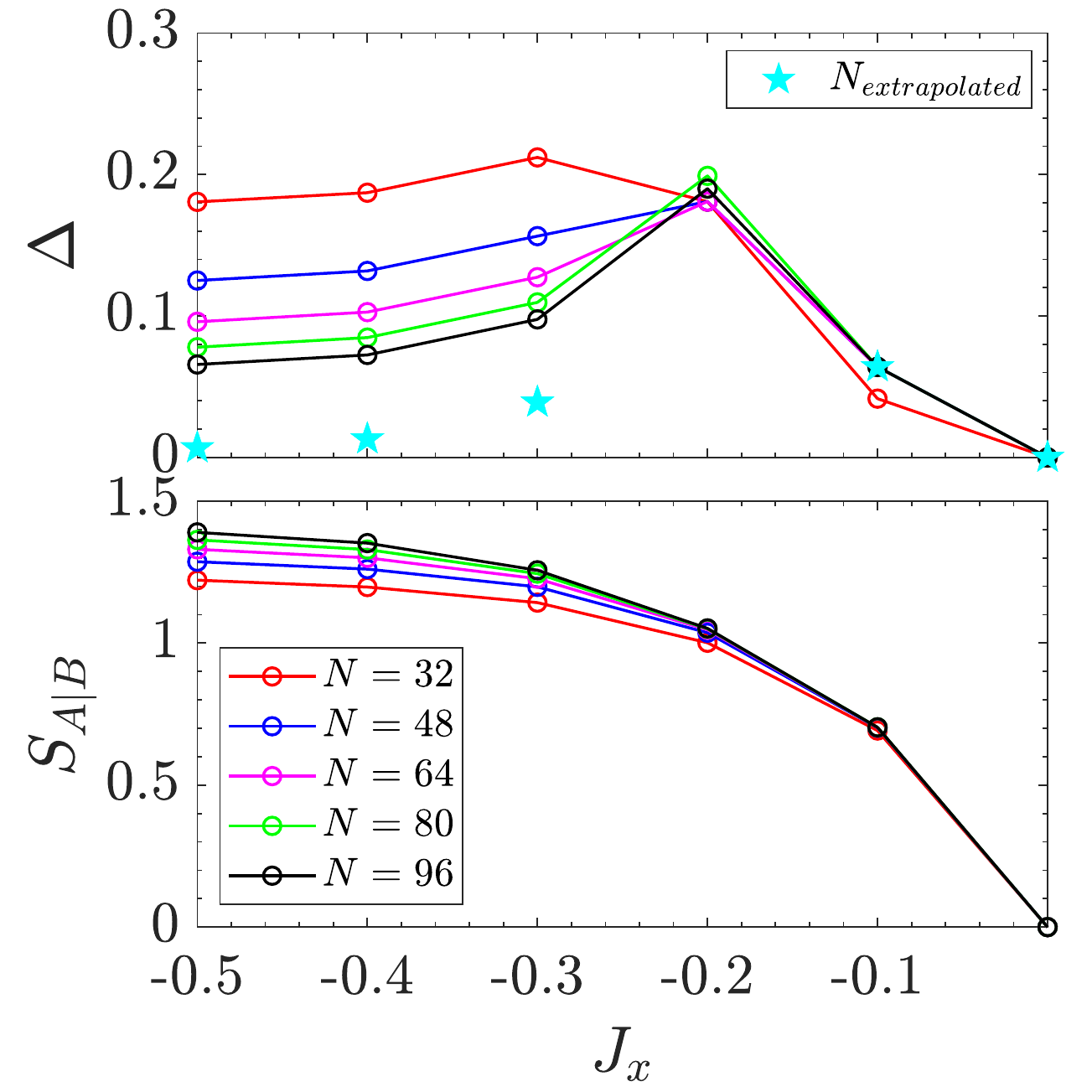}
\caption{Energy gap $\Delta$ and entanglement entropy $S_{A|B}$ vs. $J_x$ for $J_z>0$ and $J_x < 0$.
In the power-law XY FM phase $\left|J_x\right| > 0.2$, the extrapolated energy gap closes and the entanglement entropy grows logarithmically with size. The energy gap is finite and nearly size independent in the spin-liquid phase.
}
\label{fig:gap_entanglement_negativeJx}
\end{figure}

The lower panel in Fig.~\ref{fig:structure_factor_negativeJx} shows the logarithm of spin-spin correlation $\ln(S_xS_x(r))$ as a function of the logarithm of the distance $\ln(r)$. For smaller $J_x$ the plot curves down showing exponential decay. For larger $J_x$, the plot settles down to a linear relationship indicating a power law decay of spin-spin correlations, 
as expected for a quasi-1D lattice. We note that in 3D, there is a transition to a long-range ferromagnetic XY phase at $J_x \approx -0.104$ \cite{Kato2015}. 

The energy gap $\Delta$ and bipartite entanglement entropy $S_{A|B}$ as a function of $J_x$ are shown in the upper and lower panels of Fig.~\ref{fig:gap_entanglement_negativeJx} respectively. When $\left|J_x\right|<0.2$, the energy gap is only weakly size dependent, while for $\left|J_x\right|>0.2$, the gap decreases with lattice size and clearly extrapolates to zero for $\left|J_x\right|>0.3$ as expected in a power-law XY ferromagnetic phase. 
The robust energy gap in the spin-liquid phase is in contrast to the gapless photon mode in the 3-dimensional system.
In the power-law ferromagnetic phase, the bipartite entanglement entropy, when the system is divided into two equal halves, grows logarithmically with system size and is consistent with a conformally invariant behavior with central charge $c=1$ (more details can be found in the Appendix).
These results are as expected for the power-law ferromagnetic phase in one-dimension.

\subsection{Antiferromagnetic $J_z$ and Antiferromagnetic $J_{x}$} 
\begin{figure}[hbpt]
\includegraphics[width=1\columnwidth]{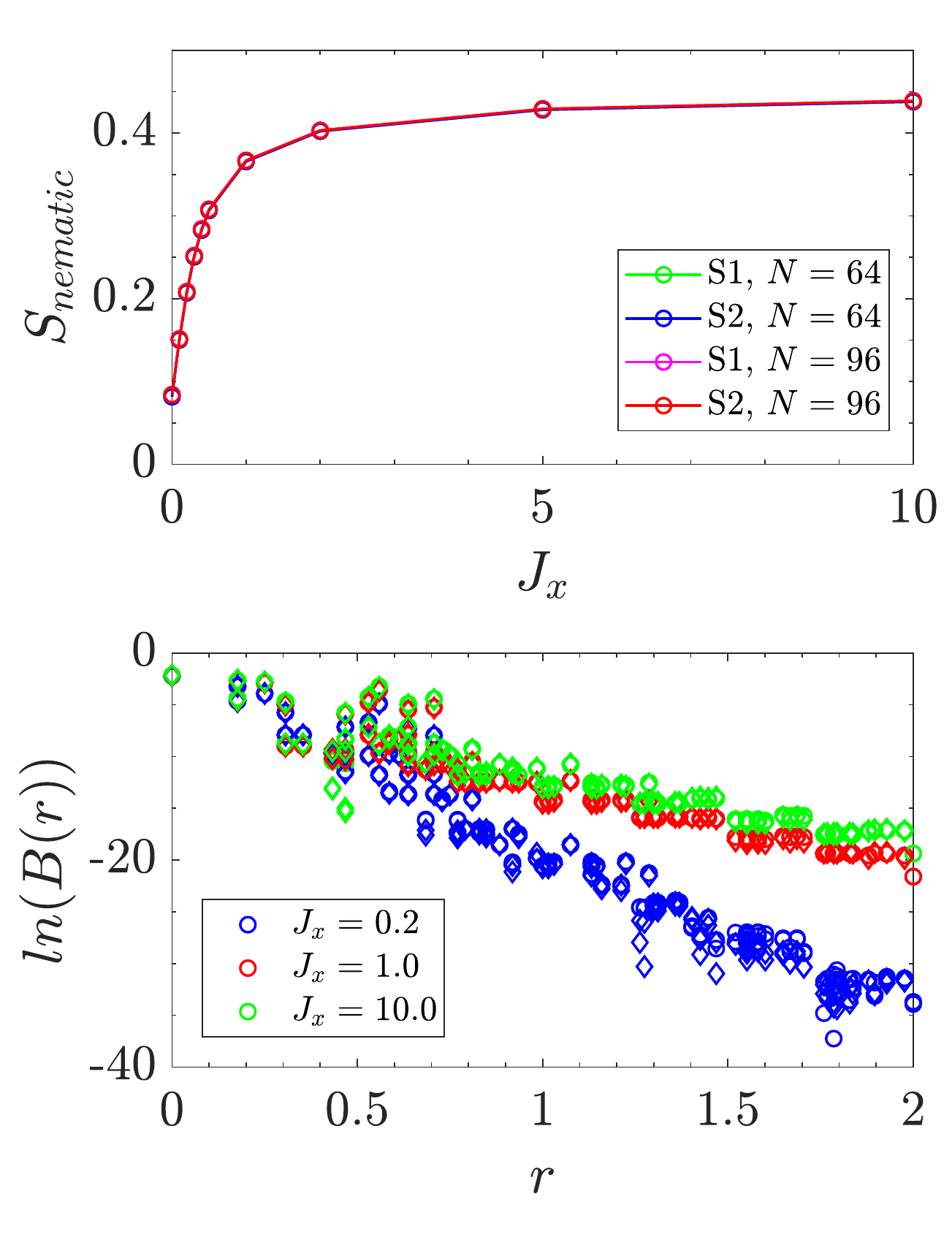}
\caption{Upper: Nematic structure factor $S_{nematic}$ as a function of coupling strength $J_x$ for $N=64$ and $N=96$ systems. Overlap of different lattice sizes curves imply that it is a short-ranged order. Lower: the logarithm of bond-bond nematic correlations as a function of the logarithm of bond distance.The slopes are $-13, -7.9, -6.8$ for $J_x=0.2, 1.0, 10.0$, indicating correlation lengths are $0.08, 0.13, 0.15$ in units of a cubic unit cell respectively for these cases. Correlation length  in terms of nearest-neighbor distance are 0.21, 0.36 and 0.42.
$N=96$ and $N=64$ data are represented by circle and diamond symbols respectively.}
\label{fig:structure_factor_positiveJx}
\end{figure}

For positive (antiferromagnetic) $J_x$ and $J_z$ couplings, we have measured spin-spin, dimer-dimer as well as nematic correlation functions. All the correlation functions show rapid exponential decay. We focus on the nematic structure factor $S_{nematic}$ defined as 
\begin{align}
S_{1}=\frac{1}{N_b} \sum_{\langle ij \rangle \langle i'j' \rangle} \langle (S_i^xS_j^x-S_i^yS_j^y)(S_{i'}^xS_{j'}^x-S_{i'}^yS_{j'}^y) \rangle
\end{align}
and 
\begin{align}
    S_{2}=\frac{1}{N_b} \sum_{\langle ij \rangle \langle i'j' \rangle} \langle (S_i^xS_j^y+S_i^yS_j^x)(S_{i'}^xS_{j'}^y+S_{i'}^yS_{j'}^x) \rangle,
\end{align}
where $\langle ij \rangle$ and $\langle i'j' \rangle$ 
denote the bonds of the pyrochlore lattice. These two types of definitions are equivalent and the results given by $S_1$ and $S_2$ are found to be identical and shown in Fig.~\ref{fig:structure_factor_positiveJx}. A long range order is not observed but a short range nematic order develops near the Heisenberg limit and persists for $J_x \gtrsim 1$. We define the corresponding nematic correlation function between two bonds with sites $(i,j)$ and $(i'j')$ respectively, as 
$B(r)= \langle (S_i^xS_j^x-S_i^yS_j^y)(S_{i'}^xS_{j'}^x-S_{i'}^yS_{j'}^y) \rangle$, where
$r$ is the distance between the bond centers.
The lower panel shows that nematic correlations decay rapidly with distance for all values of $J_x$. If one assumes a power-law decay, the power will be very large (of order 10). It is consistent with an exponential decay. 

\begin{figure}[hbpt]
\includegraphics[width=1\columnwidth]{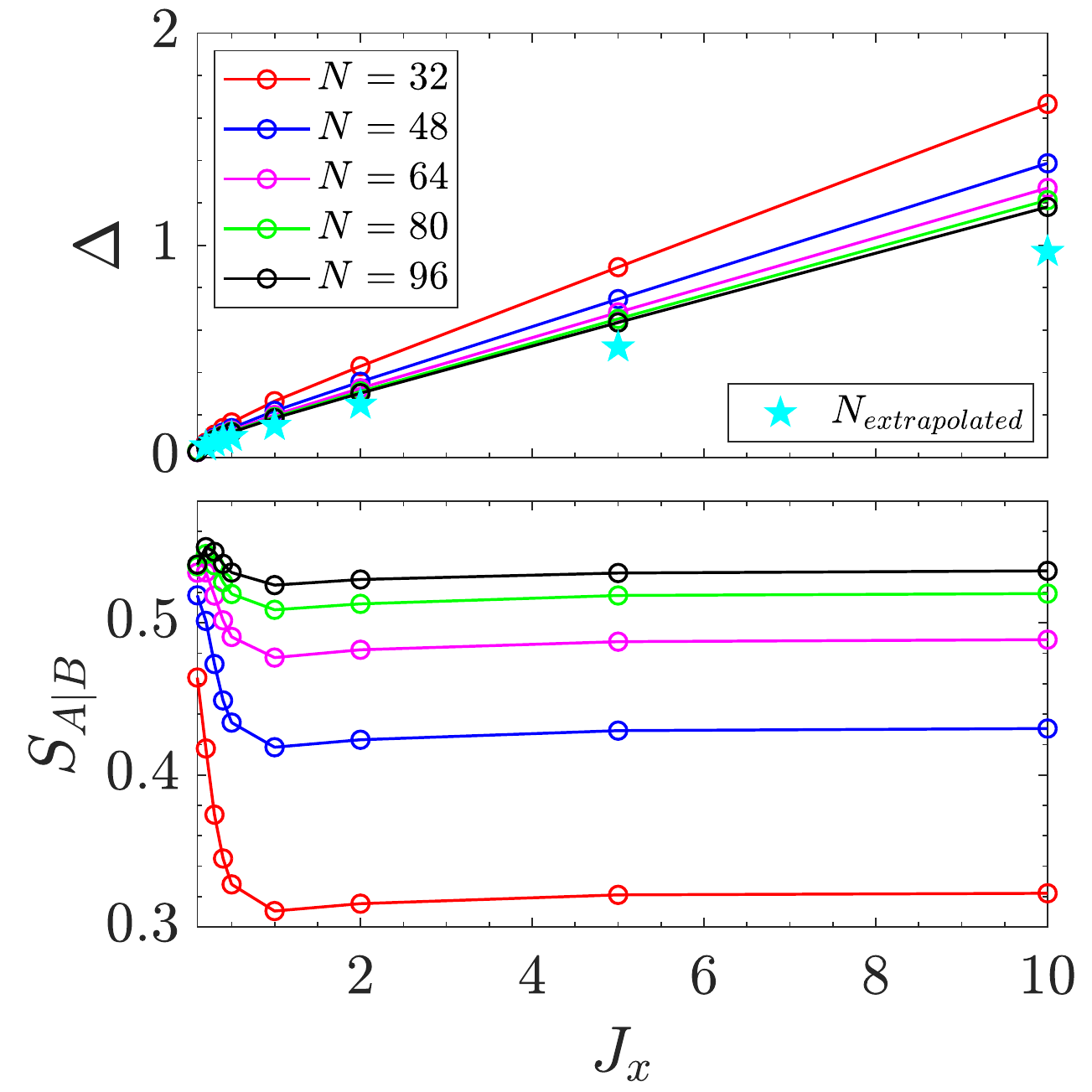}
\caption{Energy gap $\Delta$ and entanglement entropy $S_{A|B}$ v.s. $J_x$ for $J_x>0$ and $J_z>0$. A robust energy gap is observed for all $J_x$ values and entanglement entropy saturates ultimately when lattice size increases.}
\label{fig:gap_entanglement_positiveJx}
\end{figure}

Fig.~\ref{fig:gap_entanglement_positiveJx} shows the energy gap and entanglement entropy of the system for various sizes. There is a robust energy gap for all values of $J_x$. Although the entanglement entropy increases with size of the system, it ultimately saturates rather than continuing to increase. These results are consistent with a gapped ground state with a finite correlation length and no broken symmetries. 

\subsection{Ferromagnetic $J_z$ and Antiferromagnetic $J_{x}$: instability of the ferromagnetic state} 

In this section our primary goal is to study the instability of the Ising ferromagnetic state as the transverse coupling $J_x$ is increased relative to $J_z < 0$. As before we set $|J_z|=1$. Since, the z-couplings are unfrustrated, one needs to go well past $J_x=1$ to see the transition away from the fully polarized ferromagnetic state. One-particle excitations around the ferromagnetic state can be calculated analytically. Since, the Hilbert space for two-particle states in an $N$-site cluster is only of size $N(N-1)/2$, relatively large system sizes can be diagonalized exactly. 

\begin{figure}[hbpt]
\includegraphics[width=1\columnwidth]{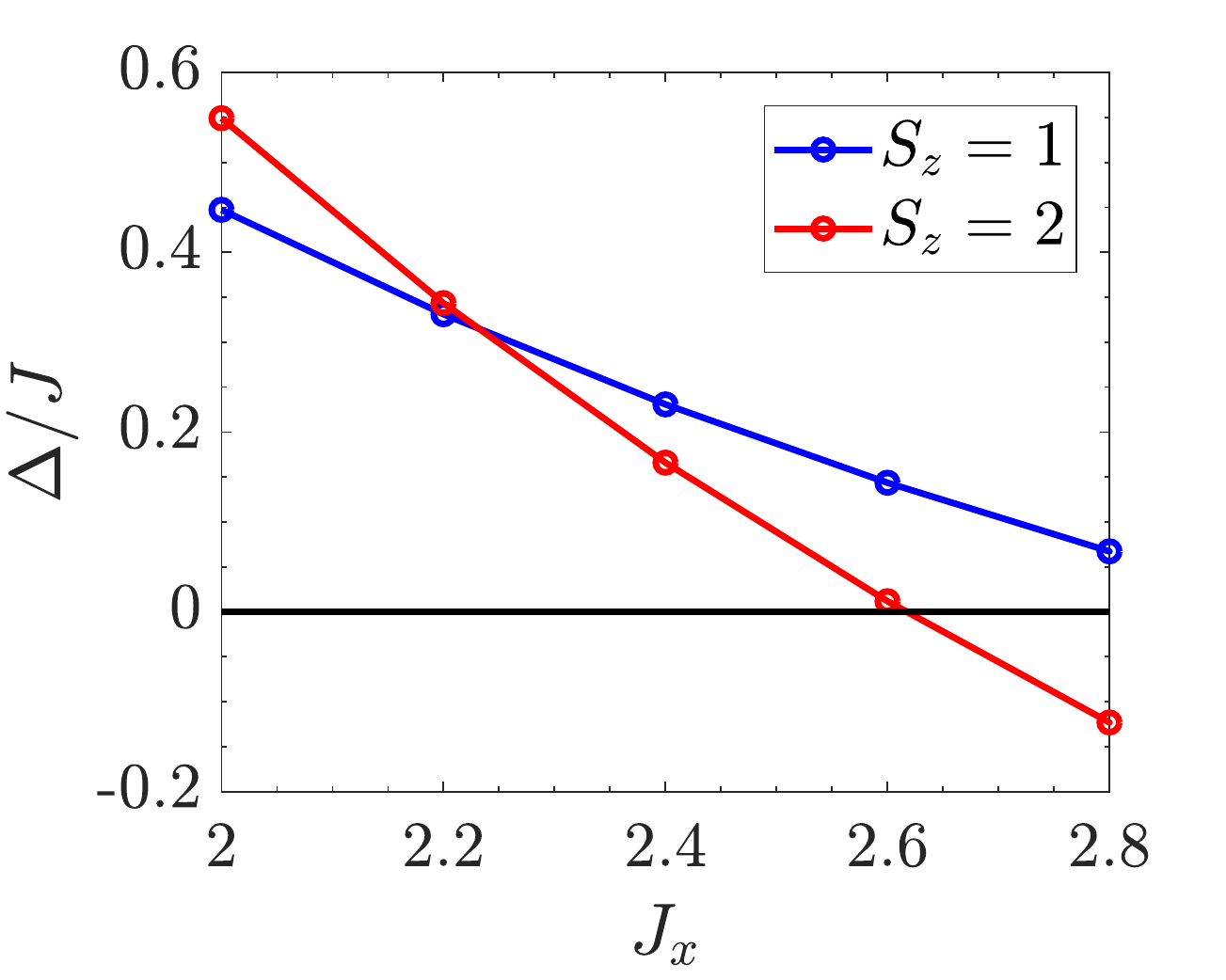}
\caption{Excitation energy $\Delta$ for $S_z=1$ and $S_z=2$ excitations from the ferromagnetic state for $J_z<0$ and $J_x>0$. The results are essentially size independent and were checked up to 128-site periodic cluster.}
\label{fig:Delta-FM}
\end{figure}

We study periodic clusters up to $N=128$ spins. The results for $S_z=1$ and $S_z=2$ excitations with respect to the ferromagnetic ground state are nearly size independent. As shown in Fig.~\ref{fig:Delta-FM}, we find that the spin-two excitations become lower in energy relative to the spin-one excitations for $J_x>2.2$. For $J_x>2.6$, the two-spins flipped state is lower in energy than the ferromagnetic state.

\begin{figure}[hbpt]
\includegraphics[width=1\columnwidth]{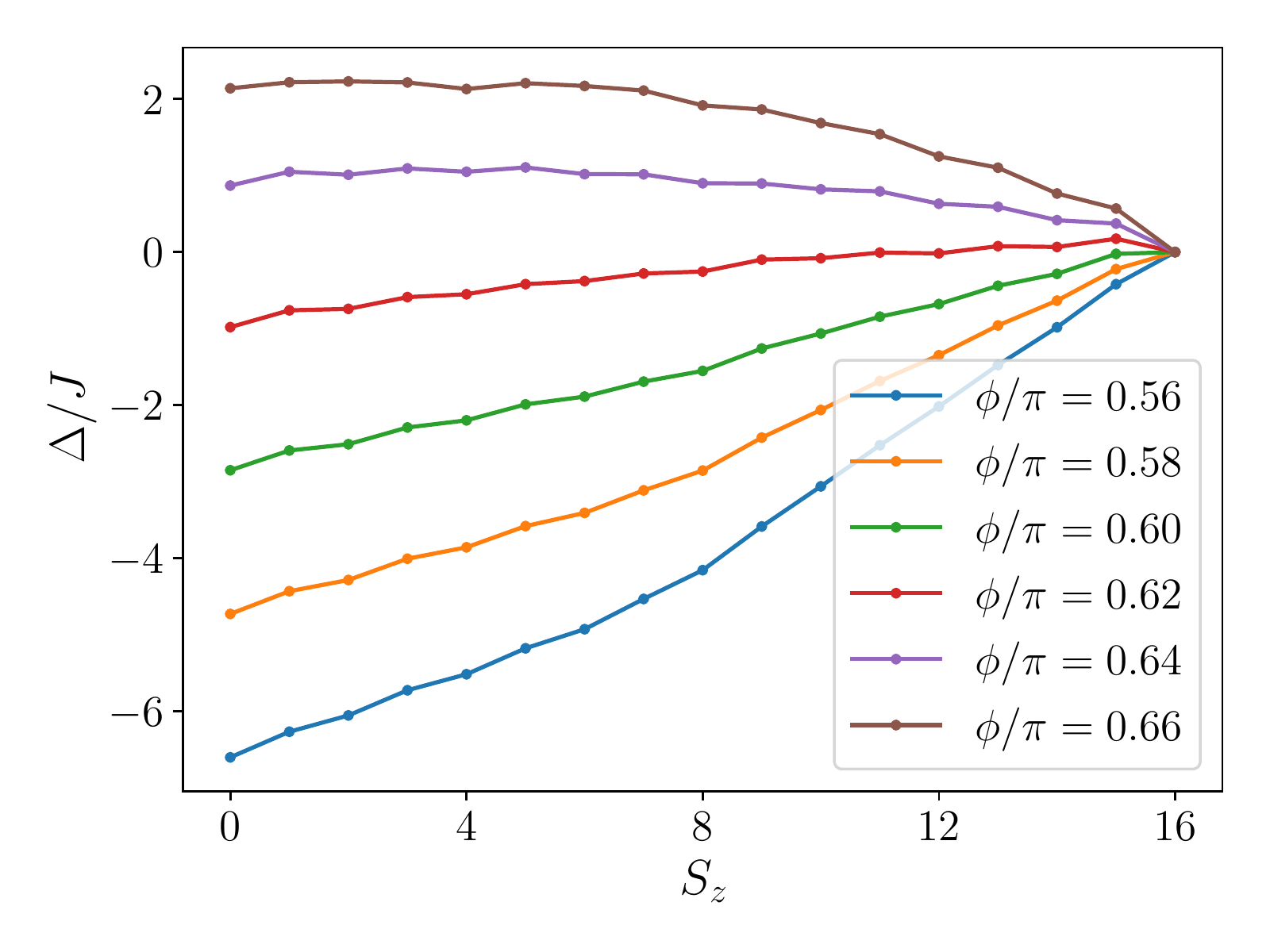}
\caption{Lowest energy states, relative to the ferromagnetic state, in different $S_z$ sectors for the 32-site periodic cluster. Close to the transition at $\phi / \pi \approx 0.61$ we observe an even-odd effect, indicative of nematic order.}
\label{fig:energy-Sz}
\end{figure}

The transition away from the ferromagnetic state happens a bit earlier. In Fig.~\ref{fig:energy-Sz} we show the lowest energy of states with different $S_z$ quantum numbers relative to the Ising ferromagnetic state for different values of $J_x$. This calculation is for the 32-site cluster. The ground state switches from the Ising ferromagnetic state to the singlet state at $J_z\approx 2.3$. As one goes into the singlet ground state, a strong odd-even behavior persists as a function of spin. These results are indicative of a nematic state, where elementary excitations have a spin-two character. 
This behavior persists as $J_x$ is further increased. Nematic states at the transition between ferromagnetic and antiferromagnetic states have been previously observed on an extended kagome lattice Heisenberg model~\cite{Wietek2020}.

One of the most noticeable aspects of the $32$-site cluster spectra in Fig.~\ref{spectrum} is the smooth behavior across the Heisenberg model point of the Hamiltonian (\mbox{$\phi=\pi/4$}). It strongly suggests that as one goes from the predominantly Ising coupling to the predominantly XY coupling, the ground state phase does not change. It argues against any special state or an increased degeneracy at the Heisenberg point. To the extent this behavior is representative of the 3D system, it argues against many proposed ground states for the Heisenberg model including dimerization, long-range nematic, broken inversion or time reversal symmetry. One possibility is that this behavior arises due to quasi-one dimensionality and the resulting confinement discussed in the next section. The short-range correlated quantum spin-liquid phase is mostly featureless.

\section{Confinement of monopole-antimonopole pairs}

In this section we examine the confinement of monopole-antimonopole pairs due to quasi one-dimensionality of the system.
To do this we perform DMRG simulations on a 192-site clusters with periodic boundary conditions in all directions. We consider the $S_z=1$ sector, which guarantees that at least one pair of monopole-antimonopole pairs must be present in each classical configuration.
We collapse many independent classical Ising configurations $\left|\psi'\right>$ from the ground state $\left|\psi_0\right>$ obtained by DMRG with probability $|\langle \psi^{'} | \psi_0 \rangle|^2$. Sampling from the desired distribution is nontrivial. An MPS based algorithm makes it realizable and a detailed discussion of the algorithm can be found in \cite{Stoudenmire10,Ferris12} The location of monopole-antimonopole pair and their distance is recorded.
In this way the probability distribution function associated with their separation is constructed and by plotting it against the logarithm of the separation, the confinement length is obtained.

\begin{figure}[hbpt]
\includegraphics[width=1.1\columnwidth]{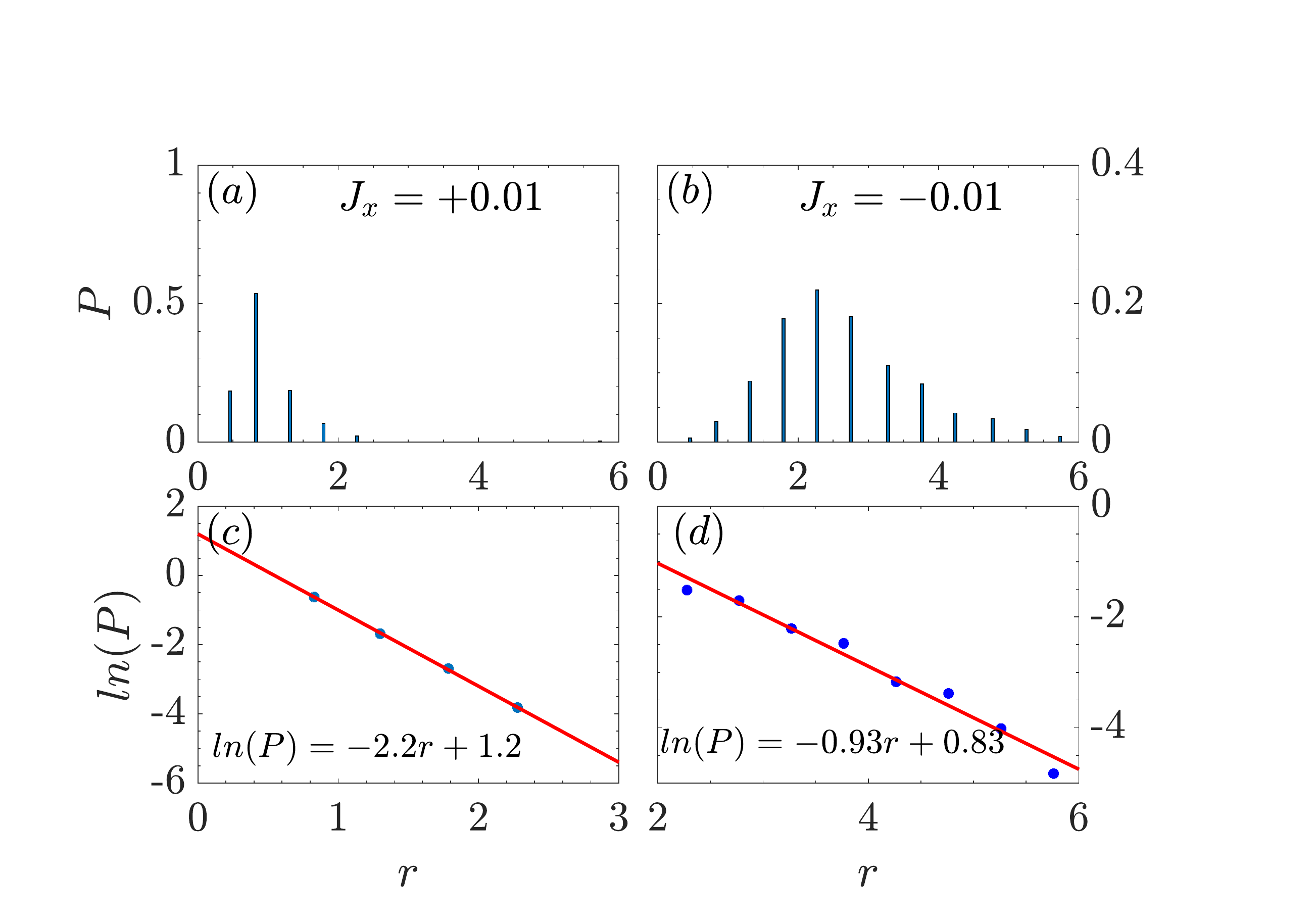}
\caption{(a)(b)Probability of monopole-antimonopole pair separations $P$ v.s. pair separations $r$ (c)(d)logarithm of the probability $\ln(P)$ v.s. separation $r$}
\label{fig:confinement}
\end{figure}

We find for antiferromagnetic $J_x$ coupling $J_x=+0.01$, the probability $P$ of separation between monopole-antimonopole pairs decreases exponentially as the separation $r$ increases. The probability almost vanishes when $r>2$. We also get a linear relationship between the logarithm of the probability $\ln(P)$ and pairs separation $r$ in the lower panels. The fitting curves are $\ln(P)=-2.2r+1.2$ for $J_x=+0.01$ and $\ln(P)=-0.93r+0.83$ for $J_x=-0.01$, suggesting that the correlation lengths are $1/2.2 \approx 0.45 $ and $1/0.93 \approx 1.08 $ in units of the length of a unit cube respectively. In other words, for ferromagnetic $J_x$ the confinement distance is longer than for antiferromagnetic $J_x$.         

\section{finite temperature properties} 
\label{sec:Results on Thermodynamics}

In this section we study the finite temperature properties of the model using the METTS method. This allows us to show that basic thermodynamic properties such as internal energy, heat capacity and entropy at intermediate and high temperatures are not much affected by the dimensionality of the system. On the other hand, the low temperature properties can be qualitatively different. We are also interested in studying the rounding of entropy plateaus due to quantum fluctuations \cite{Kato2015,Applegate12,Motome06}.

\begin{figure}[hbpt]
\includegraphics[width=1\columnwidth]{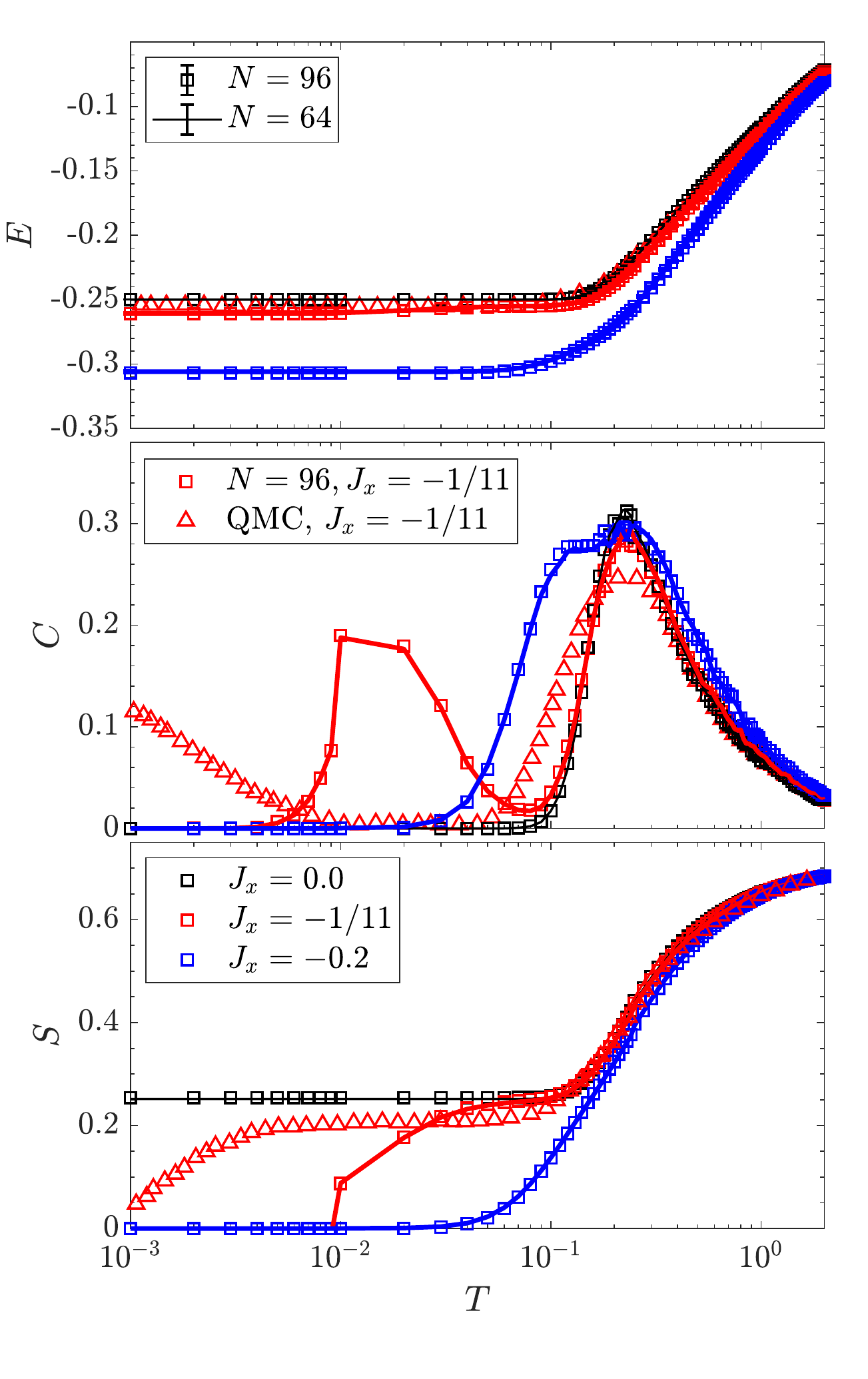}
\caption{Energy, heat capacity and entropy per site $E$ as a function of temperature $T$ for fixed $J_z=1$ and several different spin X and Y coupling strength $J_x=0,-1/11,-0.2$. Different lattice sizes $N=64$ (square point) and $N=96$ (solid lines) are shown in all the panels. The red triangular QMC data points for the 3D pyrochlore lattice are obtained from \cite{Kato2015}. 
}
\label{ECS_T_negativeJ}
\end{figure}

We first present results for an antiferromagnetic $J_z$ and a ferromagnetic $J_x$. 
We fix $J_z=1.0$ and vary $J_x$. A range of values are considered. In Fi.g.~\ref{ECS_T_negativeJ}, we show plots of the thermodynamic properties for $J_x = -1/11, -0.2$. The parameter $J_x=-1/11$ is close to the transition to the XY ferromagnetic phase in the 3D pyrochlore lattice and there is quantum Monte Carlo data available in the literature \cite{Kato2015} to compare with. In the figure, Energy $E$ as a function of temperature $T$ are shown for two different lattice sizes $N=64$(square points), $N=96$ (solid lines). Heat capacity $C$ and Entropy $S$ are obtained by using the formula $C(T)= \frac{dE}{dT}$ and $S(T)=S(T_{max})-\int _{T}^{T_{max}} \frac{C}{T}dT$ respectively.
We assume $S(T=0.01,J_x=-0.2)=0$ and $S(T=0.01,J_x=1.0)=0$ since the system is gapped at these $J_x$ values. For negative $J_x$ entropy $S(T_{max},J_x)$ is assumed to be equal to $S(T_{max},J_x=-0.2)$ and for positive $J_x$, $S(T_{max},J_x)$ is determined by $S(T_{max},J_x=1.0)$.

For $J_x=-1/11$, 
when cooling the system from high temperatures, entropy decreases from $\ln{2}$ gradually and forms a plateau near the Pauling entropy $S_p=\frac{1}{2} \ln{\frac{3}{2}} $ in the region $ 10^{-2}<T<10^{-1}$.  A comparison with previous QMC data on the 3D pyrochlore lattice (red triangle curve) is shown. At high T there is good agreement, which is consistent with the general expectation that only local coordination dictates the high-temperature behavior. However, the low temperature behavior is quite different and the drop in entropy to zero and a second heat capacity peak occurs at a much higher temperature around $T \sim 0.02$ in the quasi-1D system. By $J_x=-0.2$, there are no signs of an entropy plateau in the data and the entropy rapidly drops to zero at an even higher temperature. 

\begin{figure}[hbpt]
\includegraphics[width=1\columnwidth]{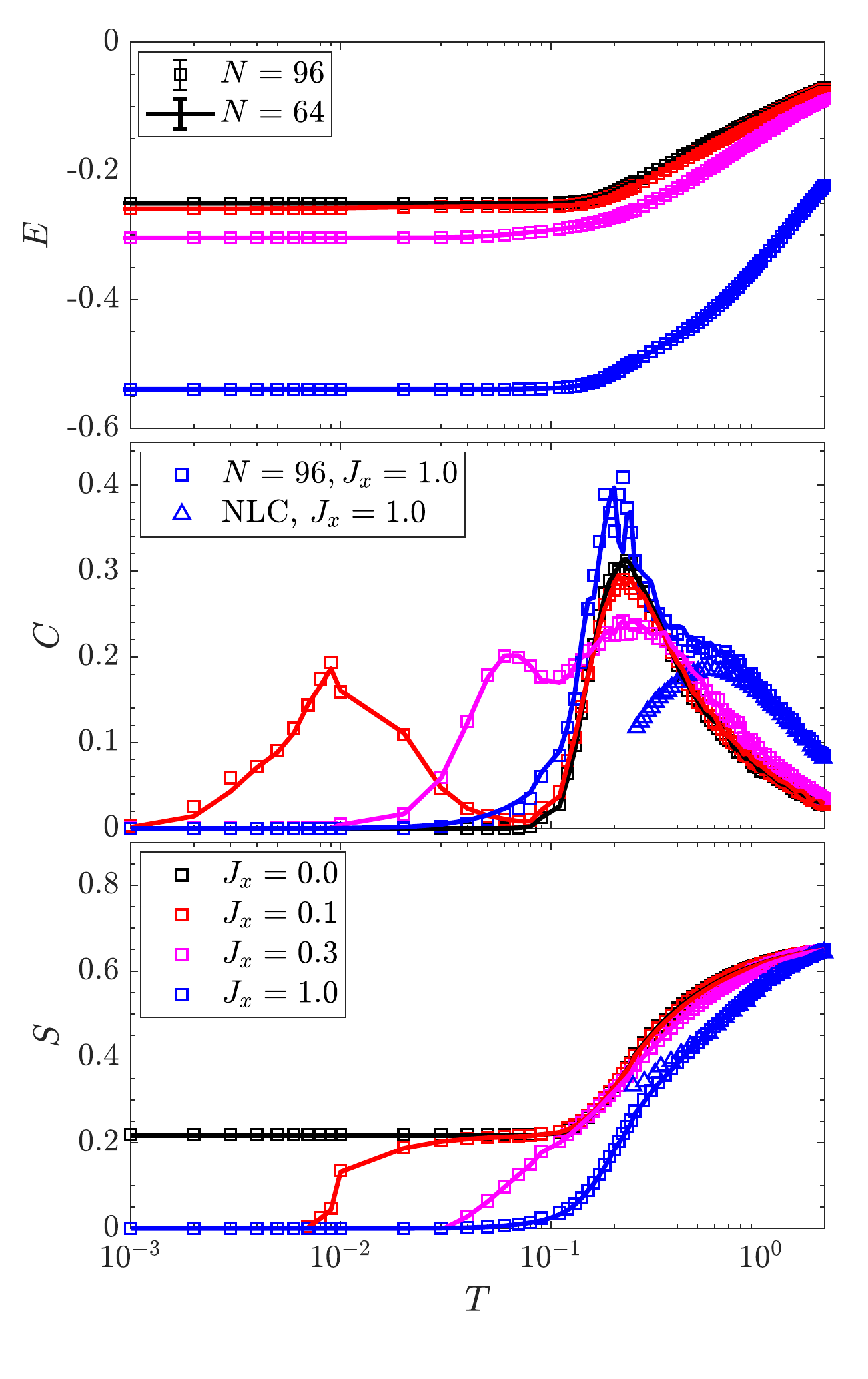}
\caption{Energy, heat capacity and entropy per site $E$ as a function of temperature $T$ for fixed $J_z=1$ and several different spin X and Y coupling strength $J_x=0, 0.1,0.3,1.0$. Different lattice sizes $N=64$ (square point) and $N=96$ (solid lines) are shown in all the panels. NLC Data \cite{Schafer20} for the Heisenberg antiferromagnet ($J_x=1.0$) for the 3D pyrochlore lattice is shown by blue triangles.}
\label{ECS_T_positiveJ}
\end{figure}

In Fig.~\ref{ECS_T_positiveJ}, the results are shown for the case of $J_x$ positive. Again, we have studied a range of $J_x$ values. We show results for $J_x=0, 0.1, 0.3$ and $1.0$. We can compare with the numerical linked cluster expansion (NLC) data \cite{Schafer20} for the 3D pyrochlore lattice in the literature at $J_x=1.0$. Once again there is very good agreement at high temperatures. However, they deviate below the peak in the NLC data at $T\approx 0.5$. 
In the quasi-1D system, the high temperature peak in the heat capacity merges with and becomes a shoulder for the low temperature peak at $T\approx 0.2 J$. However, well before $J_x=1.0$ the entropy plateau is washed out. By $J_x=0.3$, there is only a very slight hint of a shoulder in the entropy.

\section{Discussion and conclusions}

In this work we have studied the spin-half XXZ model on a pyrochlore tube. It is a model of corner sharing tetrahedra formed by joining fcc cubic cells of a pyrochlore lattice along one direction. Locally, the model has the geometry of the pyrochlore lattice and indeed high temperature thermodynamic properties are found to be in agreement with studies of the latter \cite{Kato2015,Schafer20,Benton18}. The low temperature properties are affected by the quasi-one dimensionality of the model. Our main findings are: (i) With antiferromagnetic Ising and ferromagnetic XY coupling, there is a transition from the spin ice phase to an XY ferromagnetic phase, where the latter has power-law spin correlations as expected from the dimensionality of the system. (ii) When both couplings are antiferromagnetic, there is a unique ground state with an energy gap in the system. (iii) As the parameters change from the Ising to the Heisenberg limit, short range nematic correlations develop. These correlations are nearly unchanged from the XY to the Heisenberg limit. No evidence for any long-range order of any type including nematic order is found. (iv) The low-energy spectrum evolves smoothly from the predominantly Ising to the predominantly XY couplings and the ground state of the Heisenberg model is essentially featureless. (v) For ferromagnetic Ising and antiferromagnetic XY couplings, the transition away from the fully polarized ground state is preceded by spin-two excitations becoming lower in energy than spin-one. This is further support for development of local nematic correlations in the system \cite{Benton18}. (vi) The disordered spin ice phases for both signs of the XY coupling are gapped and there is no gapless photon mode in the quasi-1D system. (vii) Monopole-antimonopole pairs are confined with a confinement length which is short, especially for antiferromagnetic transverse coupling. (viii) The plateau in entropy that exists in the purely Ising model is rounded with the addition of transverse terms. Comparison with a previous quantum Monte Carlo study of a three-dimensional system shows that the rounding is more abrupt in the quasi-1D system, where the entropy is released at a relatively higher temperature.

Computational studies of highly frustrated three-dimensional spin system remains a major challenge, especially when the system has a sign problem within quantum Monte Carlo simulations. Here we studied a simplified model making it finite in extent along two directions and much longer in the third. 
The DMRG technique can treat very long systems in one direction but only a small extent in the other two. This has allowed us to perform an unbiased study of the ground state and thermodynamic properties with high accuracy for tube geometries. One expects the short-range order found in this study, such as nematic correlations, to reflect the behavior of the full three dimensional pyrochlore system as well. Our results are consistent with a local quantum spin-liquid for the Heisenberg antiferromagnet, with only very short-range correlations and entanglement. However, the extent to which this may relate to long-range correlations/entanglement in the 3D pyrochlore lattice remains an open question that deserves further attention.

\vskip0.10in
\noindent
\underbar{\bf Acknowledgements:}
We would like to thank Owen Benton for very useful suggestions regarding studies of confinement in the model. The work of Chunhan Feng and Rajiv Singh is supported in part by NSF-DMR grant number 1855111. The Flatiron Institute is a division of the Simons Foundation.

\renewcommand{\thefigure}{S\arabic{figure}}
\setcounter{figure}{0}
\renewcommand{\thesection}{S\arabic{figure}}
\setcounter{section}{0}
\renewcommand{\theequation}{S\arabic{equation}}
\setcounter{equation}{0}

\vskip0.20in

\centerline{\large \bf Appendix}

\vskip0.10in
In this Appendix we plot the bipartite entanglement entropy $S_{A|B}$, when the system is divided into two equal halves, as a function of the logarithm of lattice size $\ln(N)$ for power-law ferromagnetic phase $J_x=-0.5, J_z=1$. The slope of the fitting curve is $c/6 \sim 0.153$, consistent with a conformally invariant behavior with central charge $c=1$.
\begin{figure}[hbpt]
 \includegraphics[width=1\columnwidth]{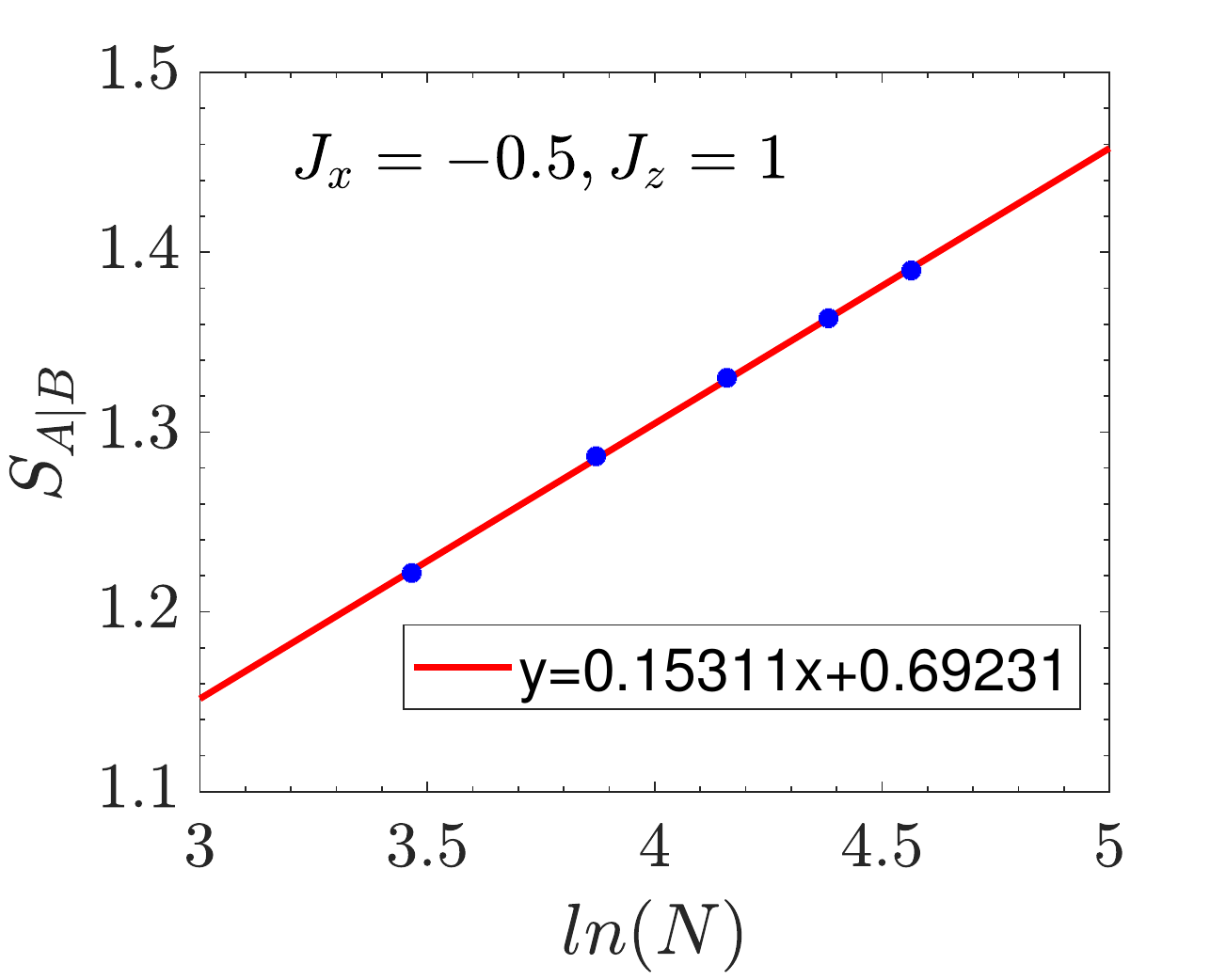}
 \caption{Bipartite entanglement entropy $S_{A|B}$ as a function of the logarithm of lattice size $\ln(N)$ for XY ferromagnetic phase $J_x=-0.5, J_z=1$. The slope of the fitting curve indicates central charge $c \sim 1$}
 \label{entanglement_entropy_lnN}
 \end{figure}
\bibliography{pyrochlore}
\end{document}